\documentclass[12pt,preprint]{aastex}
\usepackage{graphics,graphicx,rotating,amsmath}

\newcommand{\Real}[1]{{\rm Re}\left[ #1 \right]}
\newcommand{\Img }[1]{{\rm Im}\left[ #1 \right]}

\newcommand{\Pal}{{\hspace{-1pt}/\hspace{-3pt}/\hspace{-2pt}}}
\newcommand{\Ort}{{\hspace{-1pt}\times\hspace{-2pt}}}

\newcommand{\cH}{{\cal H}}
\newcommand{\bdelta}{{\boldsymbol \delta}}
\newcommand{\bq}{{\mbox{\boldmath$q$}}}
\newcommand{\bg}{{\mbox{\boldmath$g$}}}

\newcommand{\tb}{{\tilde \beta}}
\newcommand{\lr}[1]{\left( #1 \right)}
\newcommand{\slr}[1]{\left[ #1 \right]}

\begin{document}
\title{Elliptical Weighted HOLICs for Weak Lensing Shear Measurement\\
part2:PSF correction and application to Abell 370}
\author{Yuki Okura\altaffilmark{1}} 
\email{yuki.okura@nao.ac.jp}

\author{Toshifumi Futamase\altaffilmark{2}}
\email{tof@astr.tohoku.ac.jp}

\altaffiltext{1}
 {National Astronomical Observatory of Japan, Tokyo 181-8588, Japan}
\altaffiltext{2}
 {Astronomical Institute, Tohoku University, Sendai 980-8578, Japan}

\begin{abstract}
We have developed a new method(E-HOLICs) of estimating gravitational shear  
by adopting an elliptical weight function to measure background galaxy images in our previous paper.
Following to the previous paper where isotropic Point Spread Function(PSF) correction is calculated, 
we consider an anisotropic PSF correction in this paper in order to apply E-HOLICs for real data. 
A an example, E-HOLICs is applied to Subaru data of massive and compact galaxy clusters A370, and 
is able to detect double peaks in the central region of the cluster consistent with the analysis of strong lensing.
We also study the systematic error in E-HOLICs using STEP II simulation. In particular 
we consider the dependences of signal to noise ratio "S/N" 
of background galaxies in the shear estimation. 
Although E-HOLICs does improve systematic error due to the ellipticity dependence as shown in part 1, 
a systematic error due to the S/N dependence remains, namely E-HOLICs 
underestimates shear when background galaxies with low S/N objects are used. We 
discuss a possible improvement of the S/N dependence.  
\end{abstract}

\section{Introduction}

Weak gravitational lensing has been widely recognized as a unique and very powerful method to study 
not only mass distribution of the universe but also the cosmological parameters
(see for example, Mellier 1999, van Waerbeke \& Mellier 2003, Schneider 2006, Munshi et al. 2008). In particular, cosmic shear has attracted much attention recently because of its potential 
to determine the so-called cosmic equation of state, namely the relation between the effective energy density and pressure of the dark energy.
There are some detections of cosmic shear(Bacon et al 2000; Kaiser et al 2000; Maoli et al 2001; van Waerbeke et al 2001: Refregier ey al 2002; Bacon et al 2003; hamana et al 2003, Casertano et al 2003; van Waerbeke et al 2005; Massey et al 2005; Hoekstra et al 2006).
 However the lensing signal of cosmic shear is very small and thus highly accurate shear measurement is required. 
The most popular method of shear estimation is given by Kaiser et al. 1995(called the KSB method: see also Luppino \& Kaiser 1997; Hoekstra et al. 1998; Viola et al. 2011) 
where change in the moments of galactic light distribution by lensing is extracted from the measurement. 
Other methods of shear estimation have been also developed (Bernstein \& Jarvis 2002; Refregier 2003; Kuijken et al. 2006; Miller et al. 2007; Kitching et al. 2008; Melchior 2011). In relation with our E-HOLICs method, we mention the work by Bernstein \& Jarvis 2002 and Melchior 2011 who 
also introduced an elliptical weight for the shape measurement. 
All of them are very challenging and successful for cluster lensing. Although some of them have already been used for cosmic shear, it is argued that none 
of them have achieved the required level of accuracy to measure 
the cosmic equation of state in a few percent level. 

On the other hand there are planned cosmic shear observations in very near future for the purpose of measuring the cosmic equation of state. In this situation, it is urgent to develop sufficiently accurate method of shear estimation. In our previous paper(Okura \& Futamase 2011, we call part 1) we have developed a new method of the evaluating shear which is based on KSB method by 
adopting an elliptical weight function to measure the shape of background galaxies. 
Our method is a natural development of our previous studies of weak lensing analysis which uses higher order multiple moments of shape of background galaxies. 
We called the method as HOLICs(Higher Order Lensing Image Characteristics) which are quantitates with a definite spin properties made of higher-order multipole moments(Okura et al. 2007).
We have shown that Oct-HOLICs is an unbiased measure of Flexion and is successfully applied to a galaxy cluster Abell 1689 to reveal substructure in the central part of the cluster(Okura et al. 2008) and
spin-2 HOLICs which is spin-2 combination of higher order moments increases the lensing information of image and is a useful method to apply cosmic shear measurement(Okura \& Futamase 2009).
We called our method as generalized as elliptical weighted HOLICs(E-HOLICs) method because of the elliptical weight. We have showed that HOLICs method with an elliptical weight is able to measures the lensing distortion 
more accurately by weighting highly to brighter region of image than in the standard KSB method,  
and thus it can reduce effects of systematic error and random noise more effectively than KSB method.
We have also calculated isotropic PSF correction in part 1 and tested its accuracy using STEP 2 simulation data. Following to part 1, we present a method of correcting PSF containing anisotropic part 
in this paper. This is necessary to apply the method to real data. In the case of isotropic PSF correction, it was shown that the corrected shear can be obtained by solving a polynomial equation.  However, since anisotropic part has an arbitrary direction, 
it is natural to expect that the shear after PSF correction is obtained by solving coupled two polynomial equations. We succeed in reducing these coupled equations into tractable forms by dividing anisotropic part of PSF into parallel part and orthogonal part against the direction of shear(see section 2 in detail).
 We test E-HOLICs method with STEP2 simulation data.
In part 1, we can only test the data with isotropic PSF,
but in this paper we test all PSF set and obtain more detailed results.
After establishing PSF correction, we are able to apply our method to real data of galaxy cluster A370.

The organization of the paper is as follows. 
First we give a brief review of weak lensing and E-HOLICs method in section 2. 
Then in section 3, we describe the correction of PSF with anisotropic part in 
E-HOLICs method by dividing anisotropic part of PSF into anisotropic part into parallel part and orthogonal part.
In section 4 we tested E-HOLICs method by using STEP2 simulation data, in particular we investigate 
the dependences of S/N and size of background galaxies in the shear estimation.   
Then we apply E-HOLICs method to real data of galaxy cluster A 370. It will be shown that E-HOLICs 
is able to detect two peaks in the central part of the cluster consistent with strong lensing analysis. 
Finally we summarize our results and give some discussion in section 6.
\section{Bases of Weak Lensing and Elliptical weighted HOLICs}
In this section, we present bases and definitions for E-HOLICs method.
More details can be found in part 1.
\subsection{Notations and Definitions}
\begin{figure*}
\epsscale{0.5}
\plotone{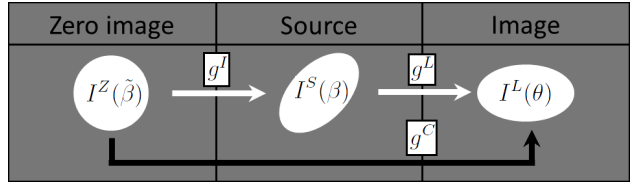}
\caption{
\label{fig:ZEROREL}
Relations between zero image, source and image.
Black arrow shows calculation in E-HOLICs methods}
\end{figure*}

Here, we describe briefly our notation and the concept of Zero image. 
The relation between zero image, source and observed image will be shown in Fig.\ref{fig:ZEROREL}. 
The bases of Weak Lensing can be found, for example, in Bartellman and Schneider 2001. 

We use complex notation for angular positions 
(e.g. $\theta=\theta_1+i\theta_2$ in the image plane and $\beta=\beta_1+i\beta_2$ in the source plane). 

For notational simplicity, we set the centroid of image as the origin in our coordinates  
(therefore centroid position is $\bar \theta=\bar \theta_1+i\bar \theta_2=0$ and $\bar \beta=\bar \beta_1+i\bar \beta_2=0$). 
We introduce an imaginary plane called zero plane where shapes of all sources are perfect circles, and regard the intrinsic shear as the result of an imaginary lensing distortion. 
The sources in the zero plane are called zero images.  
This plane is introduced in order to define naturally elliptical window in measuring shapes of background galaxy images. 
Then the lens equation gives the following relation between the displacement in zero plane $(\tilde \beta)$ and image plane$(\theta)$ as shown in part 1. 
\begin{eqnarray}
\tb=\lr{1-\kappa}\lr{\theta - \bg^C\theta^*},
\end{eqnarray}
where $\kappa$ is a dimensionless surface mass density and $\bg^C=\bg^C_1+i\bg^C_2$ 
is the "combined shear" defined as
\begin{eqnarray}
\bg^C\equiv\frac{\bg^I+\bg^L}{1+\bg^I\bg^{L*}},
\end{eqnarray}
where $\bg^L$ is the shear induced by lensing and $\bg^I$ is the intrinsic reduced shear.
Using the combined shear, we can divide lensing shear analysis into two steps as
measuring $\bg^C$ from each background galaxies and then determining lensing shear by statistical 
averaging such that 
\begin{eqnarray}
\left< \bg^I \right>\equiv \left< \frac{\bg^C-\bg^L}{1-\bg^C\bg^{L*}} \right>.
\end{eqnarray}

Henceforth, because a purpose of E-HOLICs is measuring $\bg^C$,
we notate $\bg=\bg^C$ for simplicity.
The complex distortion is defined as
\begin{eqnarray}
\bdelta\equiv\frac{2\bg}{1+g^2},
\end{eqnarray}
and the absolute values of them are notated as
\begin{eqnarray}
g&=&|\bg|\\
\delta&=&|\bdelta|.
\end{eqnarray}

We define complex moments of brightness distribution $I(\theta)$ measured with weight function which has ellipticity $\bdelta$ as
\begin{eqnarray}
Z^N_M(I,\bdelta)\equiv\int d^2\theta I(\theta) \theta^N_M W\lr{\frac{\theta^2_0-\Real{\bdelta^*\theta^2_2}}{\sigma^2}},
\end{eqnarray}
where
\begin{eqnarray}
\theta^N_M&=&\theta^{\frac{N+M}2}\lr{\theta^*}^{\frac{N-M}2}\\
W\lr{\frac{\theta^2_0-\Real{\bdelta^*\theta^2_2}}{\sigma^2}}&=&e^{-\lr{\theta^2_0-\Real{\bdelta^*\theta^2_2}}/\sigma^2}
\end{eqnarray}
and $\sigma$ is arbitrary scale (usually typical scale of each object).
Then HOLICs is defined as
\begin{equation}
\cH^N_M(I,Z^L_K,\bdelta)\equiv\frac{Z^N_M(I,\bdelta)}{Z^L_K(I,\bdelta)}.
\end{equation}

\section{PSF correction in E-HOLICs method}
In this section, we give a formulation for the correction of PSF with anisotropy in E-HOLICs method.
Generally, PSF will be arbitrary function, 
but we assume PSF has only elliptical part. 

First we define $I^L$ as a brightness distribution of a lensed image which has ellipticity $\bdelta$ and  $I^Z$ is a brightness distribution of a zero image, so 
\begin{eqnarray}
\label{eq:ILZ}
I^Z(\tilde \beta) \equiv I^L(\theta)
\end{eqnarray}
Because zero image is defined as not having ellipticity with circular weight function, we can obtain
\begin{eqnarray}
\label{eq:H22L}
\cH^2_2(I^L,Z^2_0,\bdelta)=\bdelta\\
\label{eq:H22Z}
\cH^2_2(I^Z,Z^2_0,0)=0,
\end{eqnarray}
where let $\tilde \sigma$ is a scale of weight function in the zero plane and $\tilde \sigma$ is slightly different from that in the image plane $\sigma$, because lensing changes scale of metric.
Next, we consider $\tilde I^{iso}$ which is smeared $I^Z$ by circular PSF $\tilde P^{iso}(\tilde \beta)$, therefore we have 
the following relation
\begin{eqnarray}
\label{eq:tIiso}
\tilde I^{iso}(\tilde\beta)\equiv \int d^2\tilde \psi I^Z(\tilde \psi)\tilde P^{iso}(\tilde \beta - \tilde \psi),
\end{eqnarray}
where let $\tilde P(\theta)$ as a PSF function of $\tilde P^{iso}(\tilde\beta)$ in the image plane, so
\begin{eqnarray}
\label{eq:Piso}
\tilde P(\theta) \equiv \tilde P^{iso}(\tilde\beta)
\end{eqnarray}
and because $I^Z$ and $\tilde P^{iso}$ don't have ellipticity and this relation is same as eq.(\ref{eq:H22L}) 
and eq.(\ref{eq:H22Z}), we can obtain
\begin{eqnarray}
\label{eq:H22PL}
\cH^2_2(\tilde P,Z^2_0,\bdelta)&=&\bdelta\\
\label{eq:H22PZ}
\cH^2_2(\tilde P^{iso},Z^2_0,0)&=&0.
\end{eqnarray}
And we define $\tilde I^L$ as a lensed image from $\tilde I^{iso}(\tilde \beta)$, so 
\begin{eqnarray}
\label{eq:tIL}
\tilde I^L(\theta) \equiv\tilde I^{iso}(\tilde \beta).
\end{eqnarray}
Because $\tilde I^{iso}(\tilde \beta)$ is made by convolution of circular functions,
$\tilde I^{iso}(\tilde \beta)$ also doesn't have ellipticity,
so we obtain
\begin{eqnarray}
\label{eq:H22tL}
\cH^2_2(\tilde I^L,Z^2_0,\bdelta)=\bdelta\\
\label{eq:H22tZ}
\cH^2_2(\tilde I^{iso},Z^2_0,0)=0.
\end{eqnarray}
Finally by writing eq.(\ref{eq:tIiso}) in the image plane with eq.(\ref{eq:ILZ}), eq.(\ref{eq:Piso}) and eq.(\ref{eq:tIL}), 
we obtain 
\begin{eqnarray}
\label{eq:tIL2}
\tilde I^L(\theta)= \int d^2 \psi I^L(\psi)\tilde P(\theta - \psi).
\end{eqnarray}
These equations mean that if the image (e.g. $I^L$) and the smearing function (e.g. $\tilde P$) have same ellipticity,
the smeared image (e.g. $\tilde I^L$) also has the same ellipticity with them.
Then the $\bdelta$ can be determined by observing E-HOLICs, eq.(\ref{eq:tIL2}) using $\tilde I^L$, eq.(\ref{eq:H22tL}).
On the other hand the observed image is the smeared $I^L$ by real PSF $P(\theta)$,
\begin{eqnarray}
\label{eq:Iobs}
I^{obs}(\theta)=\int d^2\psi I^L(\psi)P(\theta-\psi).
\end{eqnarray}
Thus by finding a transformation from the observed real PSF $P(\theta)$ to $\tilde P(\theta)$,
$\bdelta$ can be determined.
This is what we are going to do in the below.

In KSB method, PSF is divided into anisotropic part and isotropic part.  
However, E-HOLICs method divide PSF into orthogonal part and parallel part against the direction of lensing distortion. 
Since the direction of PSF ellipticity is not always the same with the direction of lensing distortion, it is natural to expect that the equations of correcting PSF are two dimensional form. 
By dividing PSF into orthogonal and parallel parts, PSF correction can be written in two simple one dimensional forms as shown below.
\subsection{General form of PSF correction}
Here we demonstrate general form of anisotropic PSF correction. 

Let $\bq(\theta)$ be a part of anisotropic part (it is not necessary to be whole of anisotropic part),
$P(\theta)$ can be divided into as
\begin{eqnarray}
\label{eq:PSFdecomposition}
P(\theta) \equiv \int d^2\psi P_q(\theta-\psi)\bq(\psi).
\end{eqnarray}
This decomposition is one of the basic assumption of KSB method,  
but there have been some discussion on the validity of this decomposition (Kuijken 1999). 
More careful study may be necessary on this problem.
Therefore $I^{obs}$ is
\begin{eqnarray}
I^{obs}(\theta)&=&\int d^2\psi I^L(\psi)P(\theta-\psi)=\int d^2\psi I^L(\psi)\int d^2\phi P_q(\theta-\psi-\phi)\bq(\phi)\nonumber\\
				&\equiv&\int d^2\phi I_q(\theta-\phi)\bq(\phi),
\end{eqnarray}
where
\begin{eqnarray}
I_q(\theta) \equiv \int d^2\psi I^L(\psi)P_q(\theta-\psi).
\end{eqnarray}
Let define $f(\theta)=\theta^N_MW(\lr{\theta^2_0-\Real{\bdelta^*\theta^2_2}}/\sigma^2)$, 
and by calculating moments, we obtain
\begin{eqnarray}
\int d^2\theta f(\theta)I^{obs}(\theta)&=&\int d^2\theta f(\theta)\int d^2\psi I^L(\psi)P(\theta-\psi)\nonumber\\
				&\equiv&\int d^2\theta f(\theta)\int d^2\phi I_q(\theta-\phi)\bq(\phi)\nonumber\\
				&=&\int d^2\varphi\int d^2\phi I_q(\varphi) f(\varphi+\phi)\bq(\phi).
\end{eqnarray}
By expanding $f(\varphi+\phi)$, moments of $I^{obs}$ can be expressed by combination of moments of $I_q$ and $\bq$.
Especially, $Z^2_2$ and $Z^2_0$ are obtained as
\begin{eqnarray}
\label{eq:GPSF22}
Z^2_2(I^{obs},\bdelta) &\approx& QZ^2_2(I_q,\bdelta) \nonumber\\
						&& + Q\bq^2_2\slr{Z^0_0-\frac{1}{2\sigma^2}\lr{4Z^2_0-5\bdelta^*Z^2_2}+\frac{1}{2\sigma^4}\lr{Z^4_0-2\bdelta^*Z^4_2+\bdelta^{*2}Z^4_4}}(I_q,\bdelta)\nonumber\\
						&& + Q\bq^{2*}_2\slr{  -\frac{1}{2\sigma^2}\lr{      - \bdelta  Z^2_2}+\frac{1}{2\sigma^4}\lr{Z^4_4-2\bdelta  Z^4_2+\bdelta^{ 2}Z^4_0}}(I_q,\bdelta)
\end{eqnarray}
\begin{eqnarray}
\label{eq:GPSF20}
Z^2_0(I^{obs},\bdelta) &\approx& QZ^2_0(I_q,\bdelta) \nonumber\\
						&& + Q\bq^2_2\slr{     -\frac{1}{2\sigma^2}\lr{2Z^{2*}_2-3\bdelta^*Z^2_0}+\frac{1}{2\sigma^4}\lr{Z^{4*}_2-2\bdelta^*Z^4_0+\bdelta^{*2}Z^4_2   }}(I_q,\bdelta)\nonumber\\
						&& + Q\bq^{2*}_2\slr{  -\frac{1}{2\sigma^2}\lr{2Z^2_2   -3\bdelta  Z^2_0}+\frac{1}{2\sigma^4}\lr{Z^4_2   -2\bdelta  Z^4_0+\bdelta^{ 2}Z^{4*}_2}}(I_q,\bdelta),
\end{eqnarray}
where
\begin{eqnarray}
\bq^N_M&\equiv&\frac{\int d^2\theta \theta^N_M\bq(\theta)W\lr{\frac{\theta^2_0}{\sigma^2}}}{Q}\\
Q&\equiv&{\int d^2\theta \bq(\theta)W\lr{\frac{\theta^2_0}{\sigma^2}}}
\end{eqnarray}
and $\bq^0_0=1, \bq^N_0=0(N\neq0)$.
\subsection{Parallel and orthogonal decomposition of PSF}
\begin{figure*}
\epsscale{0.5}
\plotone{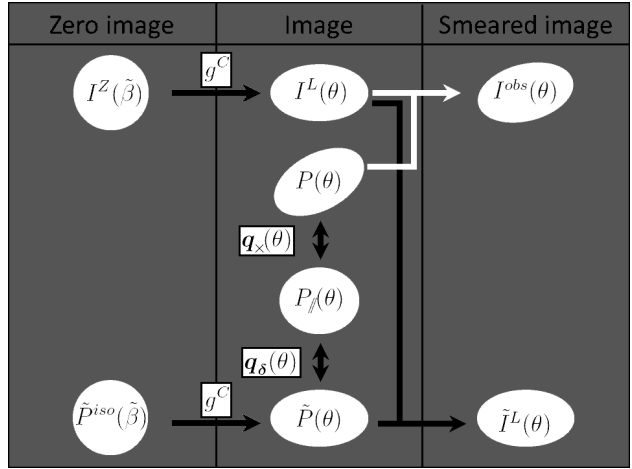}
\caption{
\label{fig:ZEROPSF}
The relations in PSF correction.
Black arrows show calculations in E-HOLICs methods}
\end{figure*}

Here, we introduce the concept of parallel part and orthogonal decomposition of PSF.
This allows us to calculate PSF correction with two simple forms.
The relations in PSF correction is shown Fig.\ref{fig:ZEROPSF}.
In KSB method, PSF is divided into isotropic PSF $P^{iso}(\theta)$ and anisotropic part $\bq^{aniso}(\theta)$ as follows. 
\begin{eqnarray}
P(\theta)\equiv \int d^2\psi P^{iso}(\psi)\bq^{aniso}(\theta-\psi),
\end{eqnarray}
Instead of the above decomposition, we decompose PSF into parallel and orthogonal part as,
\begin{eqnarray}
\label{eq: E-PSFdecomposition}
P(\theta)\equiv \int d^2\psi P_\Pal(\psi)\bq_\times(\theta-\psi).
\end{eqnarray}
where the ellipticity of $P_\Pal(\theta)$ is parallel to the ellipticity of the lensed image, 
and the ellipticity of $\bq_\times(\theta)$ is orthogonal to ellipticity of the lensed image.
The moments of these parts are defined as
\begin{eqnarray}
\label{eq:orthgonalPSF}
{\bq_\Ort}^N_M&\equiv&\frac{\int d^2\theta \theta^N_M\bq_\Ort(\theta)W\lr{\frac{\theta^2_0}{\sigma^2}}}{Q_\Ort}\\
Q_\Ort&\equiv&{\int d^2\theta \bq_\Ort(\theta)W\lr{\frac{\theta^2_0}{\sigma^2}}}
\end{eqnarray}
This decomposition might have the same problem with KSB decomposition eq.(\ref{eq:PSFdecomposition}). 
It remains to be seen if the higher order correction may correct the real PSF in a reasonably accurate level.   
 
For the moment we assume the above decomposition and define "parallel image" $I_\Pal(\theta)$ which is seared by $P_\Pal(\theta)$ as
\begin{eqnarray}
I_\Pal(\theta)\equiv \int d^2\phi I^L(\phi)P_\Pal(\theta-\phi).
\end{eqnarray}

We define $\bq_\bdelta(\theta)$ which relates $P_\Pal(\theta)$ with $\tilde P$ as follows
\begin{eqnarray}
\label{eq:PalPSF}
\tilde P(\theta)&\equiv& \int d^2\phi P_\Pal(\phi)\bq_\bdelta(\theta-\phi)
\end{eqnarray}
and moments are defined as
\begin{eqnarray}
{\bq_\bdelta}^N_M&\equiv&\frac{\int d^2\theta \theta^N_M\bq_\bdelta(\theta)W\lr{\frac{\theta^2_0}{\sigma^2}}}{Q_\bdelta}\\
Q_\bdelta&\equiv&{\int d^2\theta \bq_\bdelta(\theta)W\lr{\frac{\theta^2_0}{\sigma^2}}},
\end{eqnarray}
where ${\bq_\bdelta}^2_2$ have a same direction of $\bdelta$.
The anisotropic part $\bq^{aniso}$ in KSB decomposition has an ellipticity which has the same direction of $\bdelta$.

A catalog of these definitions can be seen in Appendix \ref{AP:def}.

For notational simplicity we set the direction of distortion as real, namely 
\begin{equation}
\bdelta=\delta
\end{equation}
Then the following expressions are obtained by definitions. 
\begin{eqnarray}
\label{eq:ortdir}
{\bq_\Ort}^2_2&=&i{q_\Ort}^2_2\\
\label{eq:qdelta}
{\bq_\delta}^2_2&=&{q_\delta}^2_2.
\end{eqnarray}

\subsection{Orthogonal PSF correction(determining a direction of distortion)}
Here, we present a method of orthogonal PSF correction.
After correcting orthogonal part, $I_\Pal$  has the direction of distortion,
therefore this is same as determining a direction of distortion.

From eq.(\ref{eq:GPSF22}), eq.(\ref{eq:GPSF20}), eq.(\ref{eq:orthgonalPSF}) and eq.(\ref{eq:ortdir}), 
the correction of orthogonal part of PSF is written as
\begin{eqnarray}
Z^2_2(I^{obs},\delta) &\approx& Q_\Ort Z^2_2(I_\Pal,\delta) \nonumber\\
						&& + {iQ_\Ort q_\Ort}^2_2\slr{Z^0_0-\frac{1}{2\sigma^2}\lr{4Z^2_0-5\delta^*Z^2_2}+\frac{1}{2\sigma^4}\lr{Z^4_0-2\delta^*Z^4_2+\delta^{*2}Z^4_4}}(I_\Pal,\delta)\nonumber\\
						&& + \lr{{iQ_\Ort q_\Ort}^2_2}^*\slr{  -\frac{1}{2\sigma^2}\lr{      - \delta  Z^2_2}+\frac{1}{2\sigma^4}\lr{Z^4_4-2\delta  Z^4_2+\delta^{ 2}Z^4_0}}(I_\Pal,\delta)\nonumber\\
						&=&Q_\Ort Z^2_2(I_\Pal,\delta) \nonumber\\
\label{eq:perpPSF22}
						&& + iZ^2_0(I_\Pal,\delta){Q_\Ort q_\Ort}^2_2\slr{H^0_0-\frac{2}{\sigma^2}\lr{H^2_0-\delta H^2_2}+\frac{1}{2\sigma^4}\lr{1+\delta^2}\lr{H^4_0-H^4_4}}(I_\Pal,Z^2_0,\delta)\nonumber\\
						&\equiv&Q_\Ort Z^2_0(I_\Pal,\delta)\Bigl(H^2_2(I_\Pal,Z^2_0,\delta) + iP^E_\Ort(I_\Pal,\delta){q_\Ort}^2_2\Bigr)\\
Z^2_0(I^{obs},\delta) &\approx& Q_\Ort Z^2_0(I_\Pal,\delta) \nonumber\\
						&& + {iQ_\Ort q_\Ort}^2_2\slr{     -\frac{1}{2\sigma^2}\lr{2Z^{2*}_2-3\delta^*Z^2_0}+\frac{1}{2\sigma^4}\lr{Z^{4*}_2-2\delta^*Z^4_0+\delta^{*2}Z^4_2   }}(I_\Pal,\delta)\nonumber\\
						&& + \lr{{iQ_\Ort q_\Ort}^2_2}^*\slr{  -\frac{1}{2\sigma^2}\lr{2Z^2_2   -3\delta  Z^2_0}+\frac{1}{2\sigma^4}\lr{Z^4_2   -2\delta  Z^4_0+\delta^{ 2}Z^{4*}_2}}(I_\Pal,\delta)\nonumber\\
\label{eq:perpPSF20}
						&=& Q_\Ort Z^2_0(I_\Pal,\delta) 
\end{eqnarray}
where we used the condition that all moments of right hand have only real parts.
Therefore we obtain 
\begin{eqnarray}
\label{eq:PSFdiv}
\cH^2_2(I^{obs},\delta) &\approx&H^2_2(I_\Pal,Z^2_0,\delta) + iP^E_\Ort(I_\Pal,\delta){q_\Ort}^2_2,
\end{eqnarray}
Thus we can divide the observed ellipticity into real part and imaginary part as follows. 
\begin{eqnarray}
{\rm Re}\slr{\cH^2_2(I^{obs},\delta)} &=&H^2_2(I_\Pal,Z^2_0,\delta)\\
{\rm Im}\slr{\cH^2_2(I^{obs},\delta)} &=&P^E_\Ort(I_\Pal,\delta){q_\Ort}^2_2.
\end{eqnarray}
Similarly the moments of star $I^{obs*}$ can be divided into as
\begin{eqnarray}
{\rm Re}\slr{\cH^2_2(I^{*obs},\delta)} &=&H^2_2(I^*_\Pal,Z^2_0,\delta)\\
{\rm Im}\slr{\cH^2_2(I^{*obs},\delta)} &=&P^E_\Ort(I^*_\Pal,\delta){q_\Ort}^2_2
\end{eqnarray}
Thus we have 
\begin{eqnarray}
\label{eq:dir}
{q_\Ort}^2_2=\frac{{\rm Im}\slr{\cH^2_2(I^{obs},\delta)}}{P^E_\Ort(I_\Pal,\delta)}=\frac{{\rm Im}\slr{\cH^2_2(I^{*obs},\delta)}}{P^E_\Ort(I^*_\Pal,\delta)}.
\end{eqnarray}
In real analysis, first we must find a direction which satisfies the second equality of eq.(\ref{eq:dir}),
and the removal of imaginary part in eq.(\ref{eq:PSFdiv}) corresponds to the correction of orthogonal PSF.

Let's consider that the observed ellipticity of image is $A+iB$ and star is $a+ib$ in arbitrary basis, respectively. 
The direction of distortion $\phi_\bdelta$ is determined by the following equation. 
\begin{eqnarray}
\phi_\bdelta=\tan^{-1}\lr{-\frac{P^E_\Ort(I^*_\Pal,\delta)B-P^E_\Ort(I_\Pal,\delta)b}{P^E_\Ort(I^*_\Pal,\delta)A-P^E_\Ort(I_\Pal,\delta)a}}.
\end{eqnarray}

\subsection{Parallel PSF correction(determining absolute value of distortion)}
Next, we show a method of correction of parallel PSF,
it is the same as correcting absolute value of the real part.

Form eq.(\ref{eq:GPSF22}), eq.(\ref{eq:GPSF20}), eq.(\ref{eq:PalPSF}) and eq.(\ref{eq:qdelta}),
we obtain the correction of parallel PSF as follows,
\begin{eqnarray}
\label{eq:palPSF}
Z^2_2(\tilde I^L,\delta) &\approx& Q_\delta Z^2_2(I_\Pal,\delta) \nonumber\\
						&& + {Q_\delta q_\bdelta}^2_2\slr{Z^0_0-\frac{1}{2\sigma^2}\lr{4Z^2_0-5\delta^*Z^2_2}+\frac{1}{2\sigma^4}\lr{Z^4_0-2\delta^*Z^4_2+\delta^{*2}Z^4_4}}(I_\Pal,\delta)\nonumber\\
						&& + {Q_\delta q_\bdelta}^{2*}_2\slr{  -\frac{1}{2\sigma^2}\lr{      - \delta  Z^2_2}+\frac{1}{2\sigma^4}\lr{Z^4_4-2\delta  Z^4_2+\delta^{ 2}Z^4_0}}(I_\Pal,\delta)\\
Z^2_0(\tilde I^L,\delta) &\approx& Q_\delta Z^2_0(I_\Pal,\delta) \nonumber\\
						&& + {Q_\delta q_\bdelta}^2_2\slr{     -\frac{1}{2\sigma^2}\lr{2Z^{2*}_2-3\delta^*Z^2_0}+\frac{1}{2\sigma^4}\lr{Z^{4*}_2-2\delta^*Z^4_0+\delta^{*2}Z^4_2   }}(I_\Pal,\delta)\nonumber\\
						&& + {Q_\delta q_\bdelta}^{2*}_2\slr{  -\frac{1}{2\sigma^2}\lr{2Z^2_2   -3\delta  Z^2_0}+\frac{1}{2\sigma^4}\lr{Z^4_2   -2\delta  Z^4_0+\delta^{ 2}Z^{4*}_2}}(I_\Pal,\delta).
\end{eqnarray}
Because these equations have only real parts, we obtain
\begin{eqnarray}
H^2_2(\tilde I^L,Z^2_0,\delta)=\delta&\approx&H^2_2(I_\Pal,Z^2_0,\delta)\nonumber\\
&&\hspace{-120pt}+{q_\bdelta}^2_2\slr{H^0_0-\frac{1}{\sigma^2}\lr{(2+3\delta^2)H^2_0-5\delta H^2_2}+\frac{1}{2\sigma^4}\lr{\lr{1+5\delta^2}H^4_0-2\delta(3+\delta^2) Z^4_2+\lr{1+\delta^2}H^4_4}}(I_\Pal,Z^2_0,\delta)\nonumber\\
		&\equiv&H^2_2(I_\Pal,Z^2_0,\delta)+P^E_\bdelta(I_\Pal,\delta){q_\bdelta}^2_2.
\end{eqnarray}
When we apply the above equation to a star, 
we define a distorted image of star $\tilde I^{*}$.
This image is obtained by using delta function for eq.(\ref{eq:tIiso}.)
Therefore $\tilde I^{*}(\theta) = \tilde P(\theta)$ and $H^2_2(\tilde I^{*},Z^2_0,\delta)=\delta$.
Thus ${q_\bdelta}^2_2$ is obtained as
\begin{eqnarray}
{q_\bdelta}^2_2=\frac{\delta-H^2_2(I^*_\Pal,Z^2_0,\delta)}{P^E_\bdelta(I^*_\Pal,\delta)}.
\end{eqnarray}
Therefore, the absolute value of the complex distortion is obtained as 
\begin{eqnarray}
\delta&=&H^2_2(I_\Pal,Z^2_0,\delta)+\frac{P^E_\bdelta(I_\Pal,\delta)}{P^E_\bdelta(I^*_\Pal,\delta)}\lr{\delta-H^2_2(I^*_\Pal,Z^2_0,\delta)}\nonumber\\
&\approx&H^2_2(I_\Pal,Z^2_0,\delta)+\frac{P^E_\bdelta(I^{obs},\delta)}{P^E_\bdelta(I^{obs*},\delta)}\lr{\delta-H^2_2(I^*_\Pal,Z^2_0,\delta)}
\end{eqnarray}
or
\begin{eqnarray}
\delta&\approx&\frac{H^2_2(I_\Pal,Z^2_0,\delta)-\frac{P^E_\bdelta(I^{obs},\delta)}{P^E_\bdelta(I^{obs*},\delta)}H^2_2(I^*_\Pal,Z^2_0,\delta)}{1-\frac{P^E_\bdelta(I^{obs},\delta)}{P^E_\bdelta(I^{obs*},\delta)}}.
\end{eqnarray}
This equation is the same as the isotropic PSF correction discussed in part 1, 
because the isotropic PSF does not have the orthogonal part.
 
Finally, we obtain the complex distortion in the two dimensional form as follows. 
\begin{eqnarray}
\bdelta&=&\cH^2_2(I_\Pal,Z^2_0,\bdelta)+\frac{P^E_\bdelta(I^{obs},\bdelta)}{P^E_\bdelta(I^{obs*},\bdelta)}\lr{\bdelta-\cH^2_2(I^*_\Pal,Z^2_0,\bdelta)}\nonumber\\
&\approx&\Real{\cH^2_2(I^{obs},Z^2_0,\bdelta)e^{-i\phi_\bdelta}}e^{i\phi_\bdelta}+\frac{P^E_\bdelta(I^{obs},\bdelta)}{P^E_\bdelta(I^{obs*},\bdelta)}\lr{\bdelta-\Real{\cH^2_2(I^{obs*},Z^2_0,\bdelta)e^{-i\phi_\bdelta}}e^{i\phi_\bdelta}},
\end{eqnarray}
where $\phi_\bdelta$ satisfies as follows
\begin{eqnarray}
\Img{\cH^2_2(I^{obs},Z^2_0,\bdelta)e^{-i\phi_\bdelta}}=\frac{P^E_\Ort(I^{obs},\bdelta)}{P^E_\Ort(I^{obs*},\bdelta)}\Img{\cH^2_2(I^{obs*},Z^2_0,\bdelta)e^{-i\phi_\bdelta}}.
\end{eqnarray}

\subsection{star selection}
In real analysis, because the ellipticity used in the weight function is obtained by PSF correction, 
we must use some technique to determine it, for example by iteration.
In each step of iteration, we must determine the centroid and measure the moments of galaxies and stars 
until the result converges,  
and thus E-HOLICs method need much longer time to measure the shear than KSB method.
If we use many stars for PSF measurement,
the time required for the analysis will increase in proportion to the number of stars.
This is not an essential problem, nut may became a practical difficulty to apply E-HOLICs for wide fields surveys.  

\section{STEP2 simulation test}
In this section, we perform tests of E-HOLICs method by using STEP2 simulation data(Massey et al 2006). 
This data set contains 6 pattern PSFs,  
and 64 types of distortion fields and the corresponding distortion fields obtained from the original  shapes in each PSF set by rotating by 90degree.
For evaluation of the shear, 1st order polynomial is used to estimate the difference between estimated shear and input shear such as
\begin{eqnarray}
\gamma^{estimated}-\gamma^{input} = m \gamma^{input} +c,
\end{eqnarray}
where $m$ means over/under-estimation, c means in/over-sufficiency of anisotropic PSF correction.
More detail information of STEP2 can be seen Massey et al 2006.
\subsection{star selection}
In real analysis, PSF varies across the field of view,
therefore we must use many stars for determining PSF distribution.
However, because STEP2 simulation data have the same PSF in the each fields,
we can correct PSF from only one star. 
To avoid the error from PSF measurement, we use only one star which has maximum S/N in each fields.

\subsection{Tests of S/N and size dependence}
A result of STEP2 test in part 1 shows that E-HOLICs method can avoid systematic error depended on intrinsic complex distortion $|\bdelta^{intrinsic}|$.
Thus we restrict ourselves to the dependencies of signal to noise ratio "S/N" and size(half light radius "rh") in the shear measurement in this paper.

\begin{figure*}
\epsscale{0.5}
\plotone{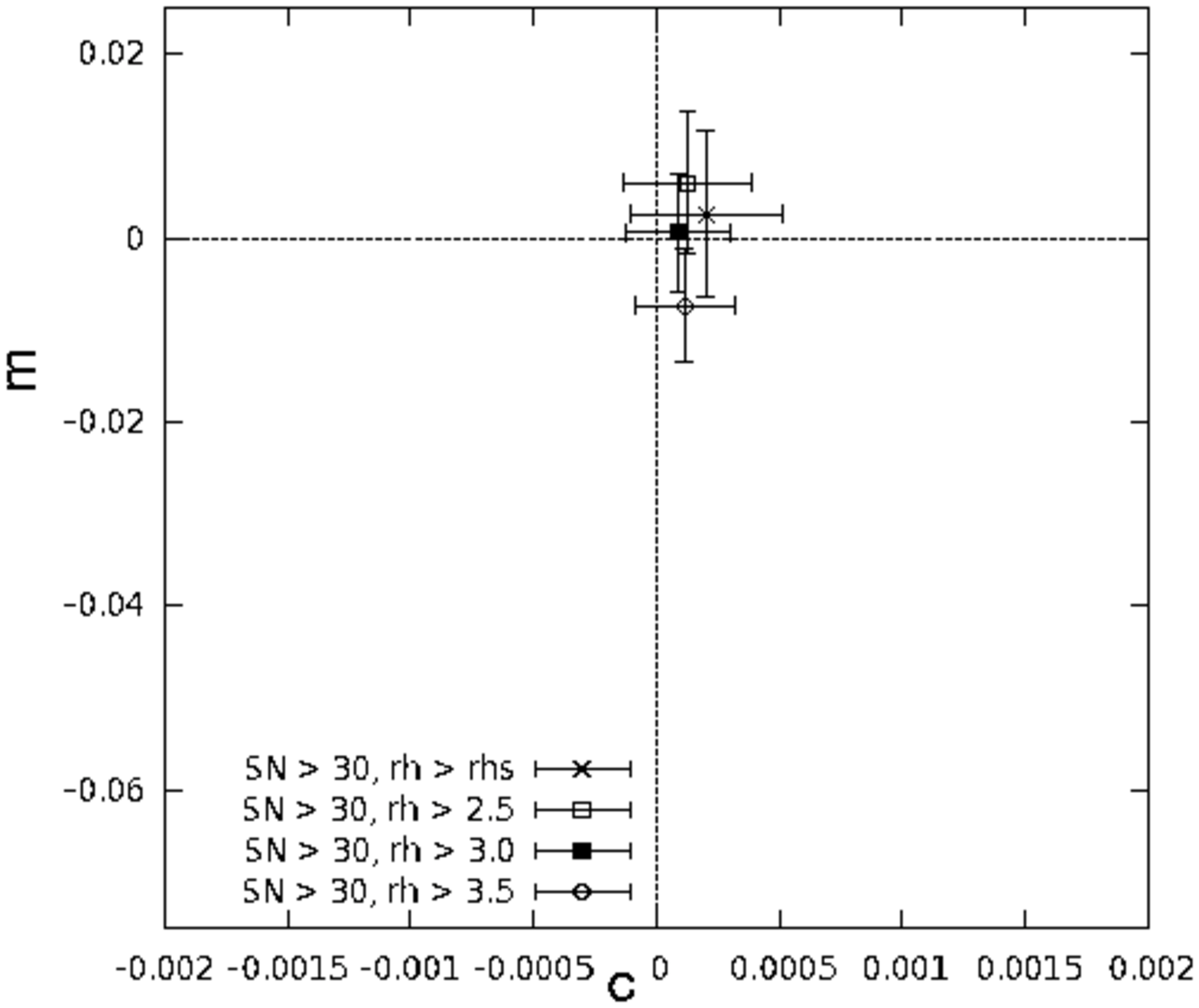}
\caption{
\label{fig:ABCDEF_rhSN30}
Result of testing ABCDEF set by using objects which are S/N$>$30 and rh$>$rhs, 2.5, 3.0 and 3.5}
\end{figure*}

\begin{figure*}
\epsscale{0.5}
\plotone{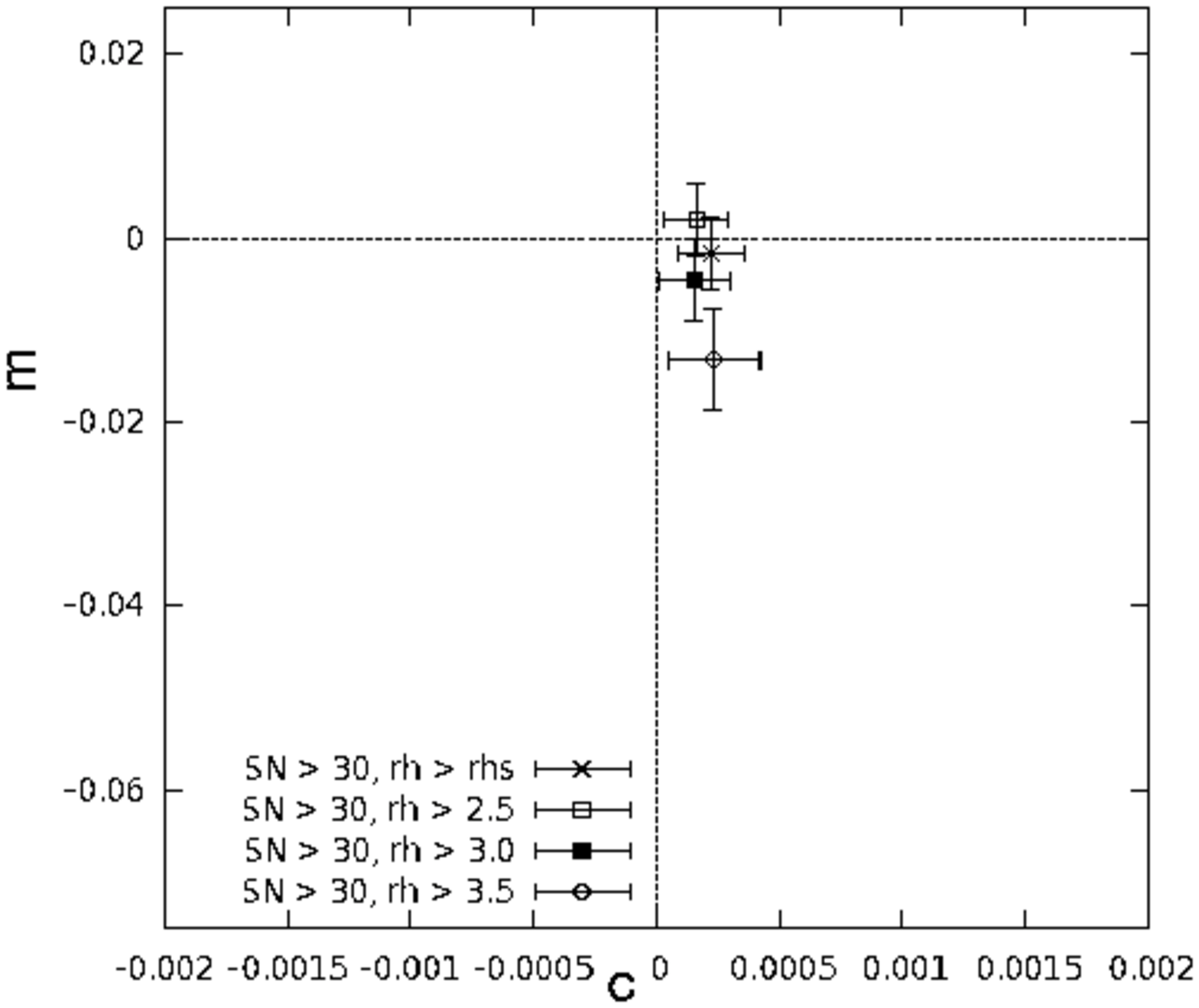}
\caption{
\label{fig:ABCF_rhSN30}
Result of testing ABCF set by using objects which are S/N$>$30 and rh$>$rhs, 2.5, 3.0 and 3.5 and}
\end{figure*}

\begin{figure*}
\epsscale{0.5}
\plotone{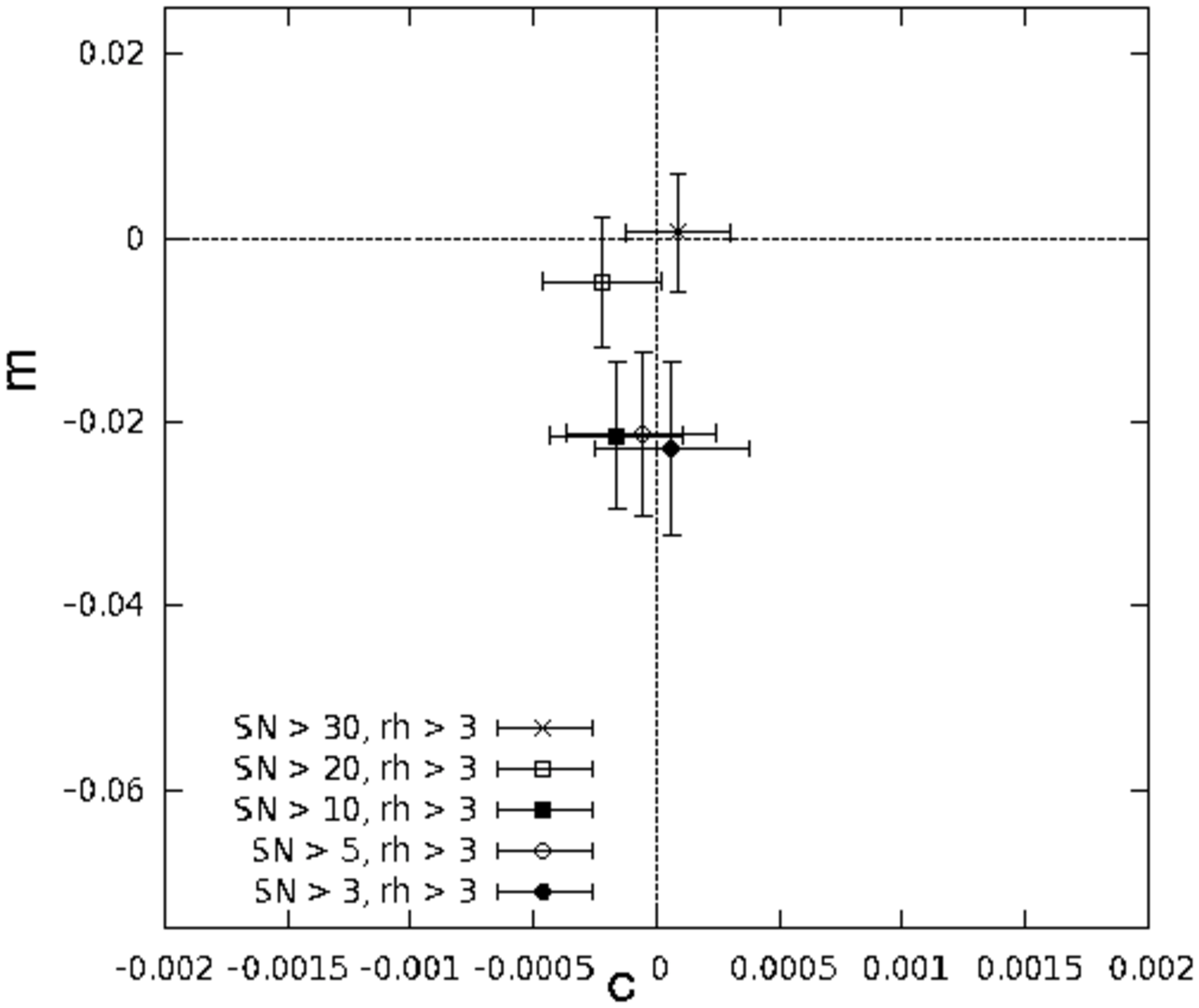}
\caption{
\label{fig:ABCDEF_SNrh3}
Result of testing ABCDEF set by using objects which are S/N$>$30, 20, 10, 5 and 3 and rh$>$3.0pix}
\end{figure*}

\begin{figure*}
\epsscale{0.5}
\plotone{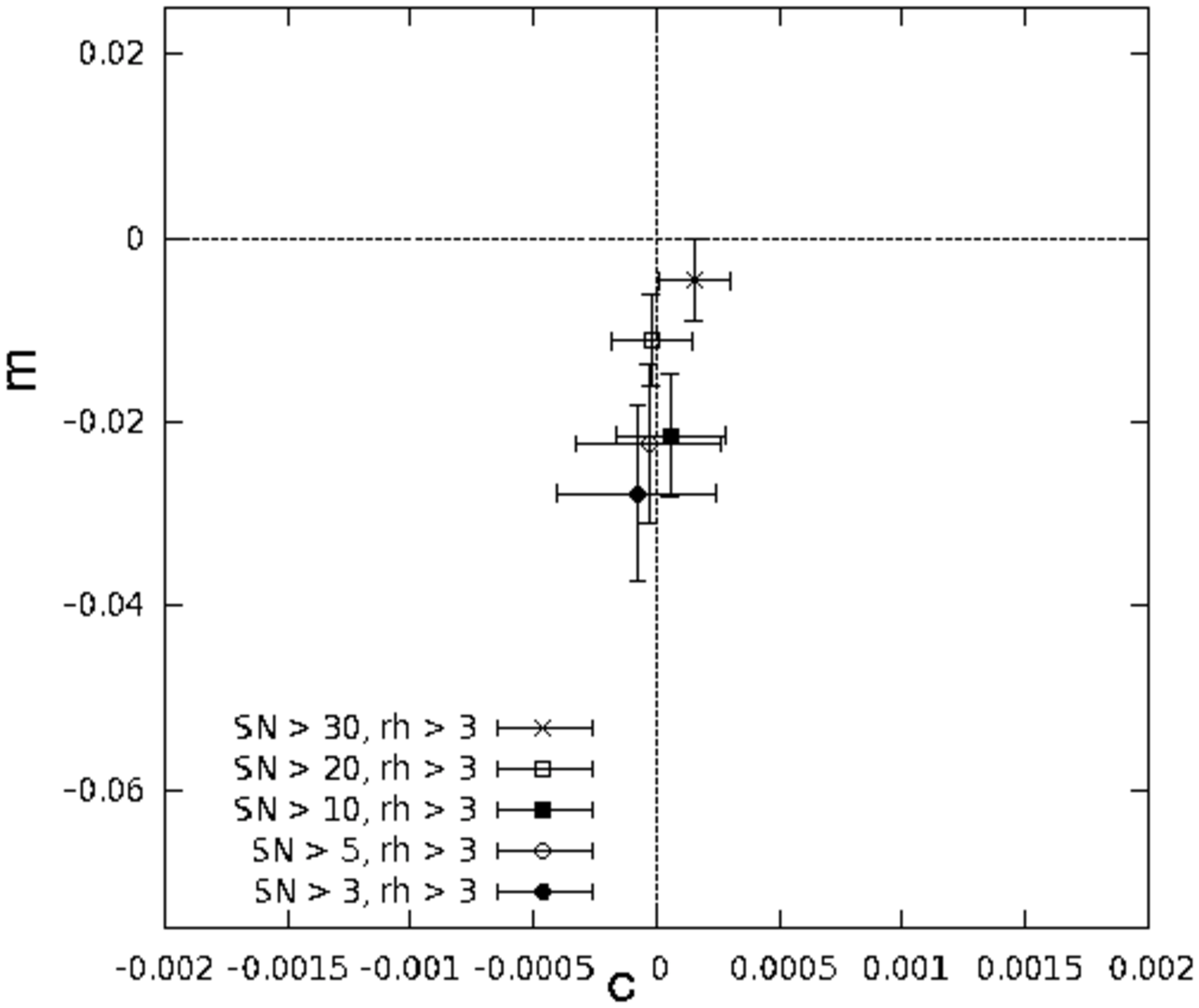}
\caption{
\label{fig:ABCF_SNrh3}
Result of testing ABCF set by using objects which are S/N$>$30, 20, 10, 5 and 3 and rh$>$3.0pix}
\end{figure*}

\begin{figure*}
\epsscale{0.5}
\plotone{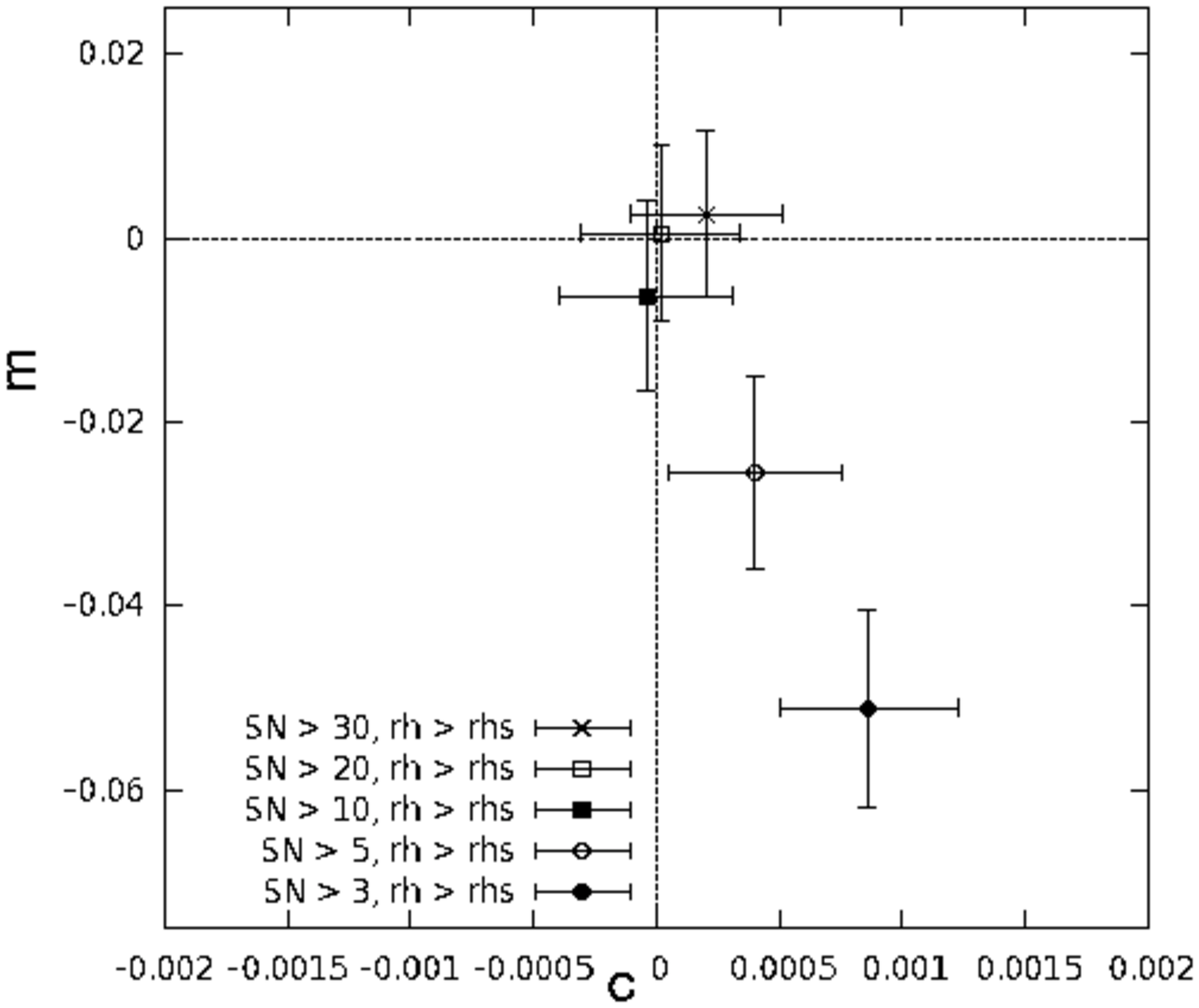}
\caption{
\label{fig:ABCDEF_SNrhs}
Result of testing ABCDEF set by using objects which are S/N$>$30, 20, 10, 5 and 3 and rh$>$rhs}
\end{figure*}

\begin{figure*}
\epsscale{0.5}
\plotone{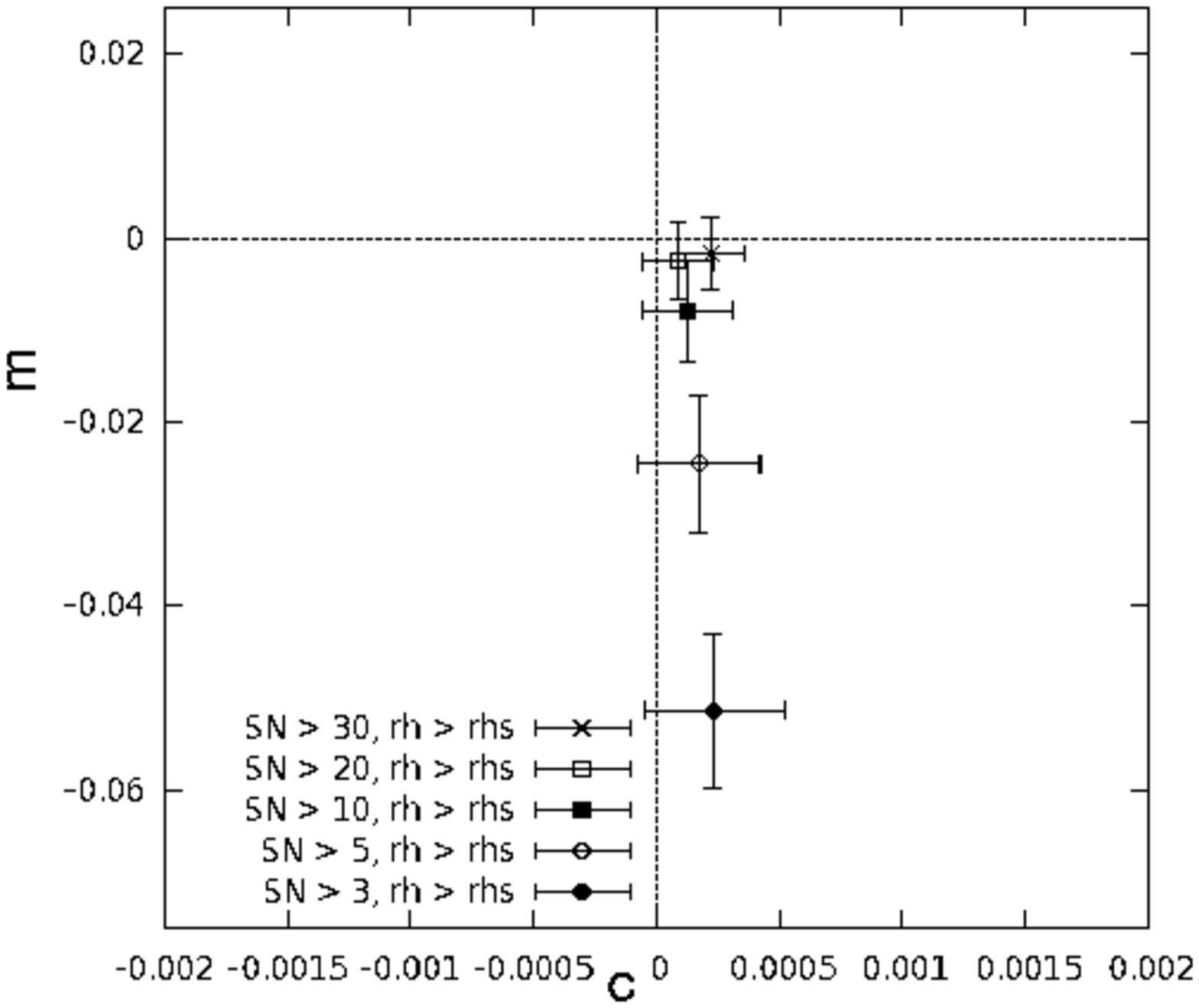}
\caption{
\label{fig:ABCF_SNrhs}
Result of testing ABCF set by using objects which are S/N$>$30, 20, 10, 5 and 3 and rh$>$rhs}
\end{figure*}

Figure \ref{fig:ABCDEF_rhSN30}. shows the results of STEP 2 test
by using objects which are S/N$>$30, rh$>$rhs, 2.5, 3.0 and 3.5 
in PSF-A, PSF-B, PSF-C, PSF-D, PSF-E and PSF-F(here and after "ABCDEF set")
Figure \ref{fig:ABCF_rhSN30} is same with Figure \ref{fig:ABCDEF_rhSN30}., but using only 
PSF-A, PSF-B, PSF-C and PSF-F(here and after "ABCF set") where "rhs" is maximum size of half light radius of star in each PSF set.  
These results show that there is a slight overestimation when small objects are used, 
but the overestimation is less than 2$\sigma$.
Figure \ref{fig:ABCDEF_SNrh3} and Figure \ref{fig:ABCF_SNrh3} are the results of STEP 2 test
by using objects which are S/N$>$30, 20, 10, 5 and 3 and rh$>$3.0 in ABCDEF set and ABCF set,  respectively.
Figure \ref{fig:ABCDEF_SNrhs} and Figure \ref{fig:ABCF_SNrhs} are the results of STEP 2 test
by using objects which are S/N$>$30, 20, 10, 5 and 3 and rh$>$rhs in ABCDEF set and ABCF set, respectively.
These results show that there is an underestimation when low S/N objects are used, particularly  
the underestimation is obvious when objects with S/N $<$ 10 are used. 

The details of some results of above selections are listed in Table 1,\\

\hspace{50pt}\begin{tabular}{|c|c|c|c|c|c|} \hline
                PSF set & S/N &    rh    & n(arcmin$^{-2}$) &      m      &       c      \\ \hline
\rule{0pt}{15pt} ABCDEF & 30 & 3.0(pix) &      2.32878      & 0.000634629 & 0.0000876483 \\ \hline
\rule{0pt}{15pt} ABCDEF & 30 &    rhs   &      3.85179      & 0.00265413  & 0.000205060  \\ \hline
\rule{0pt}{15pt} ABCDEF &  5 & 3.0(pix) &      8.38250      &-0.0212137   &-0.0000551444 \\ \hline
\rule{0pt}{15pt} ABCDEF &  5 &    rhs   &     21.9318       &-0.0255264   & 0.000403256  \\ \hline
\rule{0pt}{15pt}  ABCF  & 30 & 3.0(pix) &      2.08947      &-0.0045829   & 0.000154808  \\ \hline
\rule{0pt}{15pt}  ABCF  & 30 &    rhs   &      3.68113      &-0.00167376  & 0.000225619  \\ \hline
\rule{0pt}{15pt}  ABCF  &  5 & 3.0(pix) &      7.84198      &-0.0223220   &-0.0000304242 \\ \hline
\rule{0pt}{15pt}  ABCF  &  5 &    rhs   &     21.7678       &-0.0244922   & 0.000173477  \\ \hline
\end{tabular}\\
Table 1. The results of STEP 2 test. In this table,  S/N and rh mean the lower limit of S/N and rh of used objects, and n means number density of used objects.

\begin{figure*}
\epsscale{0.5}
\plotone{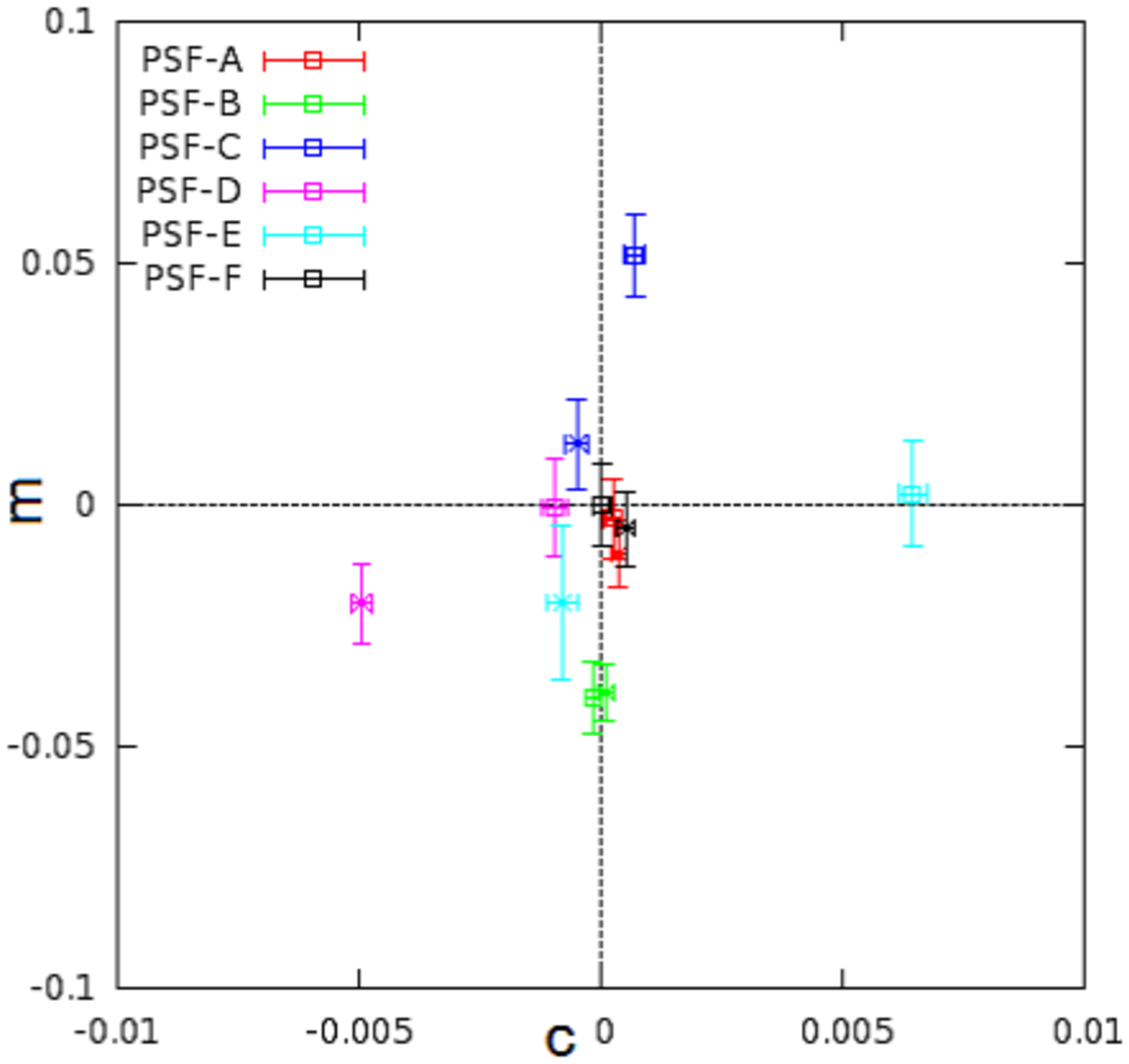}
\caption{
\label{fig:SN30rh3}
Result of each PSF set and each direction by using objects which are S/N$>$30 and rh$>$3.0pix} 
\end{figure*}

\begin{figure*}
\epsscale{0.5}
\plotone{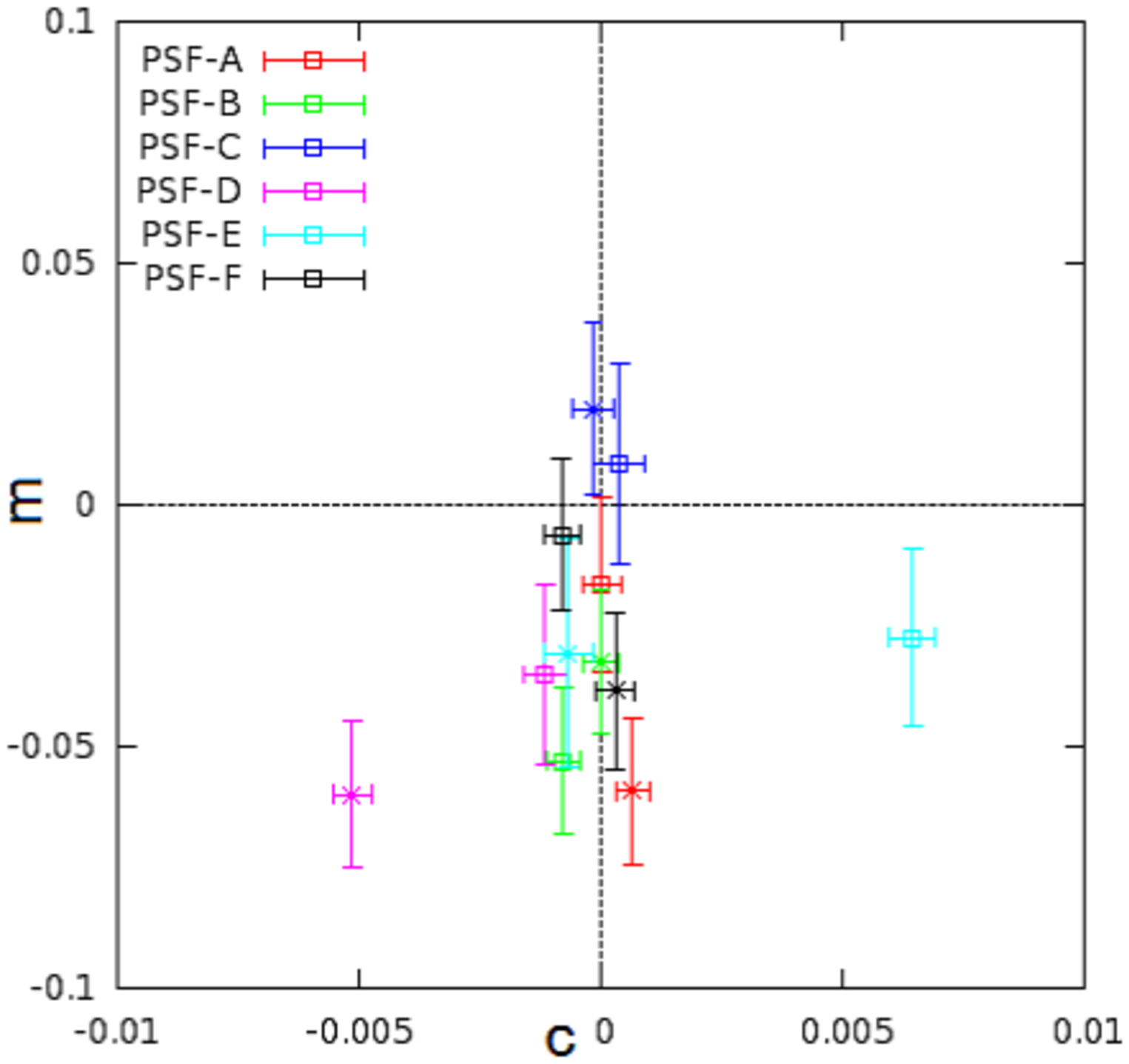}
\caption{
\label{fig:SN5rh3}
Result of each PSF set and each direction component by using objects which are S/N$>$5 and rh$>$3.0pix} 
\end{figure*}

\begin{figure*}
\epsscale{0.5}
\plotone{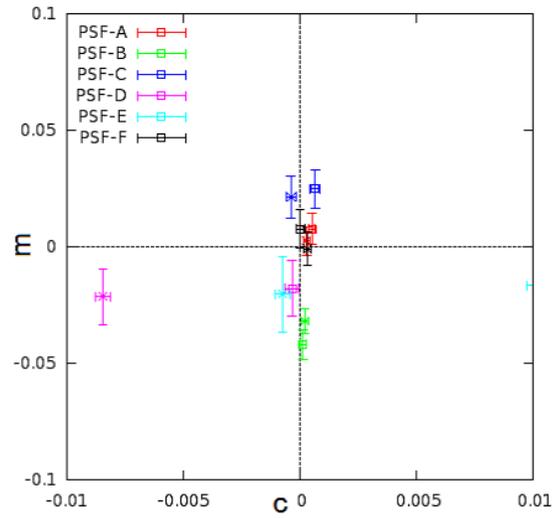}
\caption{
\label{fig:SN30rhs}
Result of each PSF set and each direction component by using objects which are S/N$>$30 and rh$>$rhs} 
\end{figure*}

\begin{figure*}
\epsscale{0.5}
\plotone{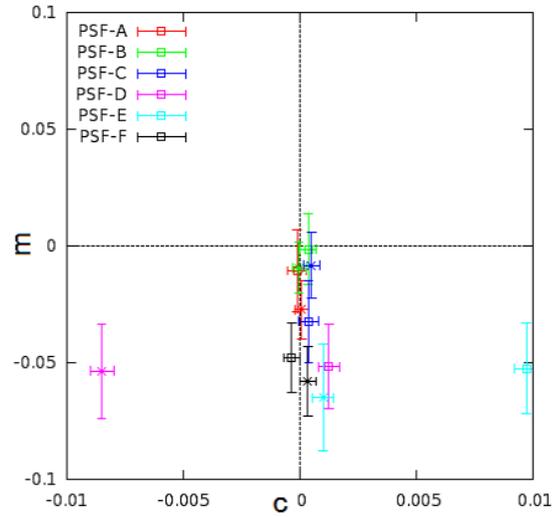}
\caption{
\label{fig:SN5rhs}
Result of each PSF set and each direction by using objects which are S/N$>$5 and rh$>$rhs} 
\end{figure*}

Next, Fig.\ref{fig:SN30rh3}, Fig.\ref{fig:SN5rh3}, Fig.\ref{fig:SN30rhs} and Fig.\ref{fig:SN5rhs}
show the results in the case of each PSF set, respectively. Here the bars with square sign 
is the result of error for the direction 1 component of the shear and the bars with cross sign 
is for the direction 2 component of the shear.The red, green, purple, pink, blue and black colors
correspond to PSF-A, PSF-B, PSF-C, PSF-D, PSF-E and PSF-F, respectively.
We can see PSF dependence from these figures,
however, there is not only PSF dependence, but also other dependences, 
and the results are affected by combination  of them.
To clearly see the effect by SN and size,
we choose above parameters.
These results show clearly the underestimation depending on S/N in all PSF set 
and the underestimation in each components in the shear can be also seen by comparing results of each PSF sets.

We can see there is an error in m in psf-B in contrast to the case of psf-A. 
Data set of psf-A and psf-B have same PSF,
but background objects have different profiles in these sets.
The profiles of background objects in psf-B are pure exponential. 
We guess that the difference in the error in m in these data set is due to the difference of radial profile 
because pure exponential is not realistic profile.
We can also see there is much larger c error for the direction 2 in PSF-D case and for the direction 1 in PSF-E case than other cases, 
but large ellipticities of input PSF are in the direction 1 in PSF-D and in the direction 2 in PSF-E, respectively.
Therefore these results look rather strange, and similar tendency is obtained by other studies
(Figure 5 of Massey et al 2007), but only "MJ" in the figure obtained good results.
Thus there is a possibility that these error may not be due to the method, but due to the data itself,
and more careful study about PSF correction may be necessary to reveal the apparent contradiction.

\subsection{Summary of STEP2 test}
From the above results of STEP 2 test, E-HOLICs method underestimates the lensing shear 
from low S/N objects, and can estimate the lensing shear precisely if we use only high S/N objects
or use an appropriate statistical weight which is function of S/N.
Systematic error depended on size is smaller than that of S/N.
This means effect of PSF is smaller than random count noise.
We do not know a fundamental reason of the S/N dependence found above. 
We need more studies for avoiding and/or correcting this systematic error for the precise determination of weak lensing shear(for example cosmic shear study).
\section{A370 galaxy cluster analysis}
In this section, we apply E-HOLICs method to the real data. 
Our data is a massive compact galaxy cluster Abell 370 at $z=0.375$ where many distorted images 
and arcs are found. 
Recently Abell 370 has been studied, for example, Richard et al. 2010 where two peaks in central region and elliptical mass distribution in outer region are found by strong lensing analysis,
Umetsu et al. 2011 and Medezinski et al. 2011 analyzed this cluster by weak lensing using Subaru telescope where 
the color information is used to make a clear separation of member galaxies from background galaxies.    
They are interested in the accurate determination of the radial mass profile of A370.
The data they used was observed with the wide-field camera Suprime-Cam (Miyazaki et al. 2002) at the prime focus of the 8.3m Subaru telescope,
and is publicly available from the Subaru archive, SMOKA5.
Subaru reduction software (SDFRED) developed by Yagi et al. 2002 was used for flat fielding, instrumental distortion correction, differential re-fraction, sky subtraction and stacking. 
More detail information of the data can be seen in Medezinski et al. 2011.
We also use the same catalogue of background galaxies with Umetsu et al. in this analysis.
Medezinski et al. 2010 and Medezinski et al. 2011 also analyzed A370 with weak lensing method.

\begin{figure*}
\epsscale{0.5}
\plotone{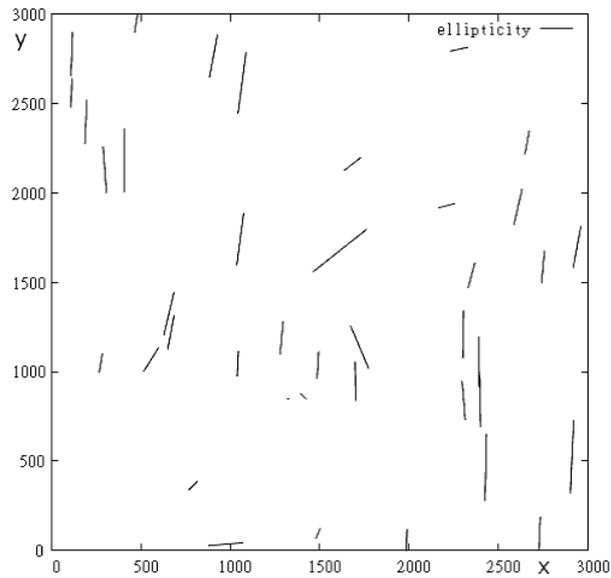}
\caption{
\label{Estar.eps}
Ellipticity of stars which are used for measuring PSF.
These ellipticity are elongated by 10000 times,
so roughly they have ellipticity 0.02 along y-axis.
} 
\end{figure*}

\begin{figure*}
\epsscale{1.0}
\plotone{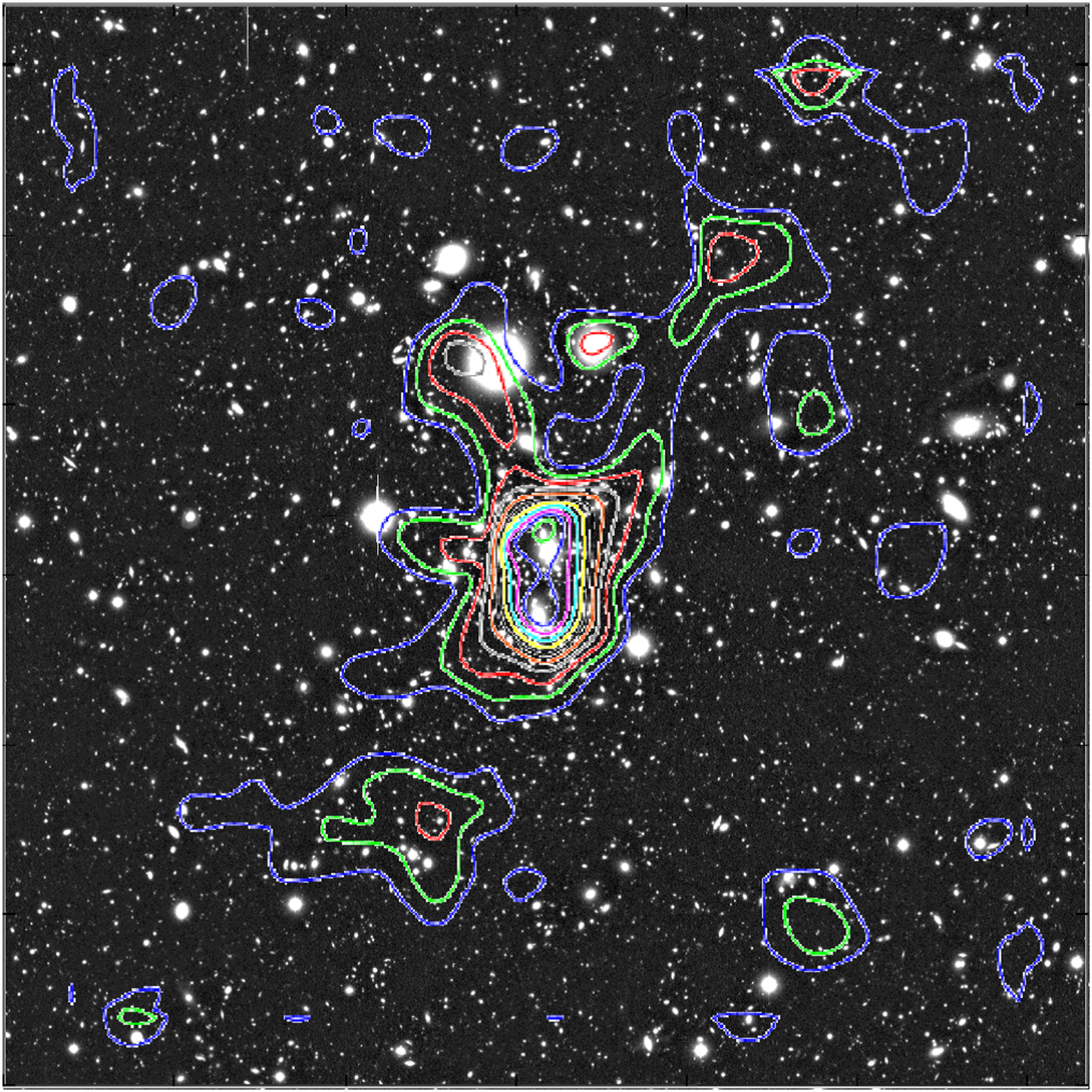}
\caption{
\label{A370z_E}
A370z analyzed by E-HOLICs method.
Contours mean $\Delta\kappa=0.1$ and minimum contour is 0.1.
} 
\end{figure*}

\begin{figure*}
\epsscale{1.0}
\plotone{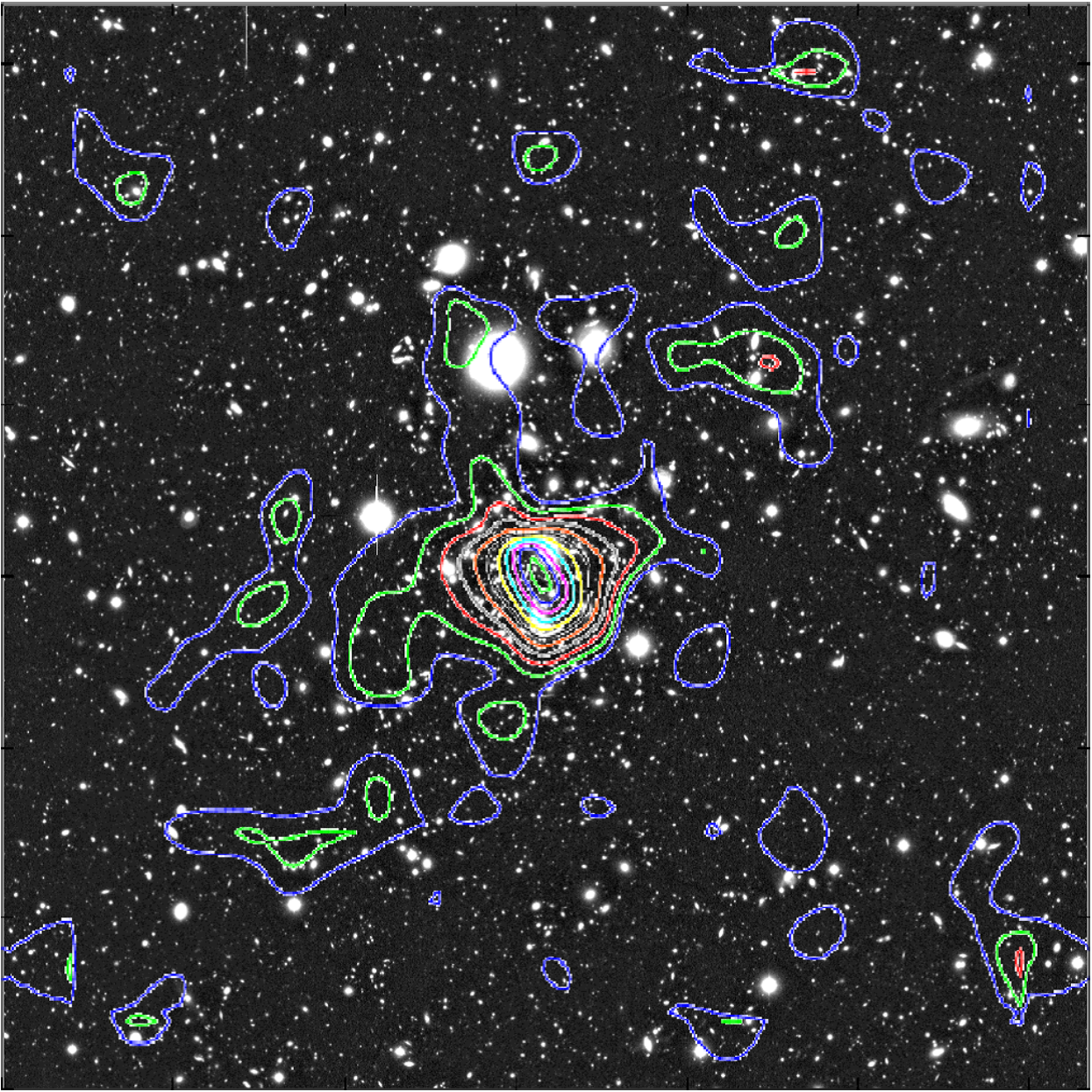}
\caption{
\label{A370z_KSB}
A370z analyzed by KSB method.
Contours mean $\Delta\kappa=0.1$ and minimum contour is 0.1.
} 
\end{figure*}

We used IMCAT(http://www.ifa.hawaii.edu/\~\ kaiser/imcat) and perl language scripts (K. Umetsu 2006, private communication) for detecting objects.
Stars for PSF correction are selected with parameters of 1.60 $<$ rh $<$ 1.95(pix), 19.0 $<$ MAG $<$ 21.0 and 4 $<$ SN,
and fig.(\ref{Estar.eps}) shows their ellipticities measured by KSB quadrupole moments.
Typically, they have ellipticities around 0.02 along y-axis.
We have used the same sample of background objects selected by using color information provided by Umetsu(Umetsu et al. 2011), 
and the parameters of rh $>$ 1.95(pix) $\approx 0.4$(arcsec) = maximum rh of stars, 25.5 $>$ MAG $>$ 20.0 and SN $>$ 5.
We used Fourier transformation(see part 1) to transform the shear distribution to the mass distribution,
where the shear distribution is smeared by Gaussian weight with 0.1arcmin scale.
Fig.\ref{A370z_E} is the results of 2 dimensional mass reconstruction of A370 by E-HOLICs method.
For comparison we show also 2 dimensional mass distribution obtained by KSB method in Fig.\ref{A370z_KSB}. 
E-HOLICs method uses 803 objects which corresponds to  7.872(arcmin$^{-2}$).
In Fig.\ref{A370z_E}, minimum contour is $\kappa=0.1$, the width between nearby contours is $\Delta\kappa=0.1$, 
and the peak value of $\kappa$ is 1.147 and S/N is 9.034 where we use RMS of B-mode as noise.
KSB method uses 776 objects which correspond to 7.067(arcmin$^{-2}$).
In Fig.\ref{A370z_KSB}, contours are similarly withdrawn as Fig.\ref{A370z_E}, but 
the peak value of $\kappa$ is 1.156 and S/N is 9.788.

We can see double peaks and elliptical mass distribution in E-HOLICs method which is consistent with the result of strong lensing(Richard 2010). On the other hand, KSB method cannot detect these peaks. 
We find that the result of KSB method has no background objects just behind A370,
but E-HOLICs method has. This will be the reason KSB method can not resolve double peaks of mass distribution.

\section{Summary and discussion}
It is widely recognized now that there are many sources of systematic errors in the evaluation of weak lensing shear and their improvement is urgently required for the precise measurement of cosmic shear. 
We have developed a new method of shear estimation, based on the KSB method and called "E-HOLICs method", by introducing the elliptical weight function to define 
multipole moments of galaxy light distribution in  previous paper(part 1). 
The use of elliptical weight function is expected to avoid the systematic error coming from an expansion of the weight function, which was usually done in some of the previous approaches to shape measurement, including the KSB method. 
Bernstein \& Jarvis 2002 and Melchior 2011 are also developing schemes which use an elliptical weight for the shape measurement. 


Following part 1 we developed the correction scheme of PSF with anisotropy in E-HOLICs method. 
Generally, the direction of the elliptical PSF is not same with the direction of distortion, and thus 
 we must treat these two direction in PSF correction,
By dividing anisotropic part of PSF into parallel and orthogonal part against the direction of the distortion, we succeeded to reduce PSF correction into two steps of simple one dimensional form. 
Then we test E-HOLICs using STEP 2 data simulation. In particular we tested the dependence on the 
S/N and size of background galaxies. 
We have found that E-HOLICs gives underestimation for low S/N sources. Although the precision is of the same level of other methods, this causes a serious problem in the accurate shear measurement. 
We have found that at least a part of reason for the underestimation is the errors in the centroid.
Although the error in the position of centroid distributes randomly, the area in the image where the measured ellipticity from the centroid within it is smaller than the correct value is larger than the other area in the image. Thus the random error in the centroid has a tendency to cause the underestimation of the measured shear.  We are investigating this effect 
and the result will be shown in the forthcoming publication.    

In this paper we have ignored higher-order moments in PSF which might be important in some cases. 
We aim to study the higher-order PSF correction also in the future publication. 
         
As we pointed out, E-HOLICs method requires longer time for measuring moments if we use many stars for PSF correction than other methods such as KSB method. This might cause a practical difficulty in any wide field surveys. 
We hope that there will be a way to avoid this problem by using powerful computer specially programed for this purpose.

As an application of E-HOLICs methods, we analyzed a massive and compact galaxy cluster Abell 370 
using by Subaru/S-Cam data.
We found that E-HOLICs can detect two mass peaks which is consistent with the strong lensing analysis. 
The KSB method cannot detect two peaks possibly because available number of background objects is reduced in compared 
with E-HOLICs. 

Although the E-HOLICs method developed in this paper has a potential to accurately measure the shear, 
it has still some shortcoming as described above and definitely more studies 
are necessary to apply E-HOLICs to the planned large-scale cosmic shear observations.

We thank K. Umetsu for a useful discussion and for providing his scripts, N. Okabe for useful discussion.
This work is supported in part by the GCOE Program "Weaving Science Web beyond Particle-Matter Hierarchy" at Tohoku University and by a Grant-in-Aid for Scientific Research from JSPS(nos. 18072001, 20540245 for TF), as well as by Core-to-Core Program "International Research Network for Dark Energy."  

\appendix
\section{List of newly defined functions}
\label{AP:def}
\hspace{-50pt}
\begin{tabular}{|c|l|l|c|} \hline
Function                             & Definition from                                                         & Ellipricity & Name \\ \hline
\rule{0pt}{15pt}$I^Z(\tb)$           &                                                                     & circular & zero image\\ \hline
\rule{0pt}{15pt}$\tilde I^{iso}(\tb)$      &$=\int d^2\tilde \phi I^Z(\tilde \phi)\tilde P^{iso}(\tb-\tilde \phi)$ & circular & smeared zero image\\ \hline
\rule{0pt}{15pt}$I^L(\theta)$        &$=I^Z(\tb)$                                                          & having a ellipticity of $\bdelta$ & lensed image \\ \hline
\rule{0pt}{15pt}$I^{obs}(\theta)$    &$=\int d^2\phi I^L(\phi)P(\theta-\phi)$                                &  & observed image\\ \hline
\rule{0pt}{15pt}$I_\Pal(\theta)$     &$=\int d^2\phi I^L(\phi)P_\Pal(\theta-\phi)$                           & having a direction of distortion & parallel image\\ \hline
\rule{0pt}{15pt}$\tilde I^L(\theta)$ &$=\int d^2\phi I^L(\phi)\tilde P(\theta-\phi)$                         & having a ellipticity of $\bdelta$ & distortional image\\ \hline
\rule{0pt}{15pt}$P(\theta)$          &$= \int d^2\phi P_\Pal(\phi)\bq_\Ort(\theta-\phi)$                    & & real PSF \\ \hline
\rule{0pt}{15pt}$\tilde P^{iso}(\tb)$&$=\tilde P(\theta)$                                                  & circular &  isotropic zero PSF \\ \hline
\rule{0pt}{15pt}$\tilde P(\theta)$   &$=\int d^2\phi P_\Pal(\phi)\bq_\bdelta(\theta-\phi)$                      & having a ellipticity of $\bdelta$ & distortional PSF \\ \hline
\rule{0pt}{15pt}$P_\Pal(\theta)$     &$=\int d^2\phi P^{iso}(\phi)\bq_\Pal(\theta-\phi)$                     & having a direction of distortion & parallel PSF \\ \hline
\rule{0pt}{15pt}$\bq_\Ort(\theta)$  &$P(\theta)=\int d^2\phi P_\Pal(\phi)\bq_\Ort(\theta-\phi)$   & orthogonal to distortion & orthogonal part of $P(\theta)$ \\ \hline
\rule{0pt}{15pt}$\bq_\bdelta (\theta)$  &$\tilde P(\theta) =\int d^2\phi P_\Pal(\phi)\bq_\bdelta(\theta-\phi)$& having a direction of distortion & distortional part of $\tilde P(\theta)$ \\ \hline
\end{tabular} 
\acknowledgments

\end{document}